\documentstyle[bezier]{article}

\newcommand{\PRL}[1]{{\sl Phys.~Rev.\ Lett.}~{\bf #1}}

\newcommand{\CQG}[1]{{\sl Class.~Quant.\ Grav.}~{\bf #1}}
\newcommand{\PRD}[1]{{\sl Phys.~Rev.}~{\bf D#1}}

\newcommand{\PR}[1]{{\sl Phys.~Rev.}~{\bf #1}}

\begin{document}

\title{{\hfill \small UMDGR 96-47, gr-qc/9511022}\\
Multi-Black-Hole Geometries in (2+1)-Dimensional Gravity}
\author{Dieter R. Brill
\thanks{e-mail: Brill@umdhep.umd.edu}\\
{\small Department of Physics, University of Maryland,
College Park, MD 20742, USA }}
\maketitle

\begin{abstract}
Generalizations of the Black Hole geometry of Ba\~nados, Teitelboim and
Zanelli (BTZ) are presented. The theory is three-dimensional vacuum Einstein
theory with a negative cosmological constant. The $n$-black-hole solution
has $n$ asymptotically anti-de Sitter ``exterior" regions that join in one
``interior" region. The geometry of each exterior region is identical to that
of a BTZ geometry; in particular, each contains a black hole horizon that
surrounds (as judged from that exterior) all the other horizons. The
interior region acts as a closed universe containing $n$ black holes. The
initial state and its time development are discussed in some detail for the
simple case when the angular momentum parameters of all the black holes
vanish. A procedure to construct $n$ black holes with angular momentum
(for $n \geq 4$) is also given.

\end{abstract}
\section{Introduction}

Since Ba\~nados, Teitelboim and Zanelli's \cite{BTZ} discovery of black holes
in 2+1 dimensional Einstein theory there has been
considerable interest in finding solutions that describe several black
holes. Such solutions exist in 3+1 dimensional general relativity under
special circumstances. The best-known solution of this type
is probably the static configuration of $n$ charged black holes
whose gravitational and electromagnetic forces balance \cite{MP}. However,
dynamic Multi-Black-Hole (MBH) solutions are known as well, for instance the
MBH cosmologies of Kastor and Traschen \cite{KT}. The essential requirement
for simple, closed-form MBH solutions appears to be absence of specific
interactions between the black holes, due to some special balance condition,
for example on the charge/mass ratio. In 2+1 dimensional Einstein theory,
where spacetime curvature is uniquely determined by the cosmological constant
$\Lambda$, there can be no specific interaction between bodies. One would
therefore expect such MBH solutions to exist in 2+1 dimensions.

For positive or vanishing $\Lambda$ the global structure of spacetime is too
rigidly determined by the curvature to admit even single black holes. For
negative $\Lambda$ the BTZ solution \cite{BTZ} has all the expected properties
of a black hole in a negative curvature, asymptotically anti-de Sitter (adS)
environment; for a comprehensive review see ref.\ \cite{C}.

In this paper we shall construct MBH generalizations of the BTZ solution.
We will find it convenient to approach the solutions first from the point of
view of the initial value problem. Existence of solutions of the initial value
constraints establishes existence of the corresponding spacetime, at least
for a finite interval in time. We will then discuss this time development
of the MBHs.

In 3+1 dimensions the constraints are easier to solve than
the full dynamics, and MBH initial values can be given for a variety of
masses, charges, and topologies (see for example refs.\ \cite{M} and
\cite{BL}). There is also a variety of horizon structures, as determined
by the apparent horizons \cite{BL,T}; for example, two black holes may
initially exhibit only the apparent horizon of each hole, or they may
have three horizons, with the extra one surrounding both holes. In 2+1
dimensions we obtain only the latter scenario with
regular initial values and asymptotically anti-de Sitter exterior regions.
This is to be expected because the exterior regions are static and therefore
cannot contain more than one regular black hole \cite{CH}.

In section 2 we recall the single BTZ black hole and show how its initial
geometry can be represented in the case of vanishing angular momentum $J$.
Section 3 gives a construction of initial data for MBHs without angular
momentum ($J=0$). The time development of
these MBHs is discussed in Section 4. Section 5 shows how to construct
MHBs with non-vanishing angular momentum.

\section{BTZ Initial Values}

In three dimensions the Ricci tensor determines algebraically the full
Riemann tensor. A three-dimensional Einstein space therefore has constant
curvature proportional to $\Lambda$. The BTZ solution is of this type and
takes the form, in Schwarzschild coordinates \cite{BTZ,C},
\begin{equation} \label{BTZ}
ds^2 = -(|\Lambda|r^2-M)dt^2 + f^{-2}dr^2 + r^2d\phi^2  - Jd\phi dt.
\end{equation}
Here $\phi$ is an angular coordinate with period $2\pi$, and
$$f^2 = \left(|\Lambda|r^2-M + {J^2\over 4r^2}\right).$$
The zeros of $f^2$ are denoted by $r_+$ and $r_-$, so that
$$M = |\Lambda|(r_+^2 + r_-^2) \qquad {\rm and}
\qquad J^2 = 4|\Lambda|r_+^2r_-^2.$$
When $J=0$ the BTZ geometry is time-symmetric about $t = {\rm const}$,
so this initial two-dimensional spacelike surface must
also have constant negative curvature. Its universal covering is therefore
two-dimensional hyperbolic space, $H$. Most of the subsequent discussion
will concern the geometry of $H$ and the identifications implied by the
periodicity of $\phi$, rather than coordinate expressions such as (\ref{BTZ}).

We shall find it convenient to use two representations of $H$.
One is its isometric embedding in Minkowski space, for example as the past
spacelike hyperboloid of constant spacetime distance from the origin.
The isometries of $H$ are the boosts and rotations of three-dimensional Lorentz
transformations. As applied to $H$, the boosts are called transvections
(whereas rotations remain rotations). Each transvection leaves one geodesic
invariant. In the embedding this geodesic is the intersection of the
hyperboloid with the plane through the origin orthogonal to the boost's axis.
Conversely, any (directed) geodesic segment determines a transvection.

The other representation is Poincar\'e's open disk.
This can be regarded as a ``stereographic" projection of the Minkowski
hyperboloid on its tangent plane, the projection center being the point
diametrically opposite to the tangent plane's point of tangency.
The map is conformal (like the stereographic projection of
the sphere), the boundary of the disk represents ideal points at infinity,
circles look like circles, and geodesics are represented by circles that
meet the boundary orthogonally. Two geodesics intersecting at infinity
are said to be parallel, and two geodesics that do not intersect are
called ultraparallel. Two ultraparallel geodesics determine a unique geodesic
segment that is normal to both and represents the shortest distance between
them. The segment's transvection moves one of the geodesics into the other.

The initial state of black holes with $J=0$ can now be described in terms of
the geometry of $H$: any two ultraparallels
represent a $t=0$ surface of a BTZ black hole, cut open along a geodesic
$\phi =$ const. Namely, the common normal to the two ultraparallels
determines a one-parameter family of transvections corresponding to
$\phi$-displacement in the BTZ metric. The BTZ initial geometry itself
corresponds to the region between the ultraparallels. (These should be
identified to reassemble the uncut BTZ geometry.) Because the $r={\rm const}$
curves of BTZ space are metrically circles, they are also (parts of) circles
on the Poincar\'e disk, and they are orthogonal to the ultraparallels. The
unique circle (or straight line) of this set that is a geodesic corresponds
to the black hole horizon. So the horizon defines the transvection by which
$\phi = 0$ and $\phi = 2\pi$ are identified. These features are illustrated
in Fig.\ 1.

\bigskip
\unitlength 1.00mm
\begin{picture}(75.00,50.00)(-10,5)
\bezier{116}(56.00,42.00)(64.00,30.00)(56.00,18.00)
\bezier{116}(24.00,18.00)(16.00,30.00)(24.00,42.00)
\thicklines
\bezier{152}(24.00,41.83)(40.00,31.17)(56.00,41.83)
\bezier{152}(24.00,18.17)(40.00,28.83)(56.00,18.17)
\thinlines
\put(40.00,36.50){\line(0,-1){13.00}}
\put(23.50,42.00){\vector(3,-1){0.2}}
\bezier{108}(9.00,32.00)(13.50,47.00)(23.50,42.00)
\put(23.50,18.00){\vector(3,1){0.2}}
\bezier{108}(9.00,28.00)(13.50,13.00)(23.50,18.00)
\put(9.00,30.00){\makebox(0,0)[cc]{identify}}
\put(57.00,42.00){\makebox(0,0)[lc]{$\phi=2\pi$}}
\put(57.00,18.00){\makebox(0,0)[lc]{$\phi=0$}}
\put(75.00,35.00){\vector(-3,-1){15.00}}
\put(77.00,35.00){\makebox(0,0)[cb]{asymptotic region}}
\bezier{20}(24.00,42.00)(30.00,50.00)(40.00,50.00)
\bezier{20}(56.00,42.00)(50.00,50.00)(40.00,50.00)
\bezier{20}(24.00,18.00)(30.00,10.00)(40.00,10.00)
\bezier{20}(56.00,18.00)(50.00,10.00)(40.00,10.00)
\put(30.00,32.00){\vector(3,-1){10.00}}
\put(30.00,33.00){\makebox(0,0)[cb]{horizon}}
\bezier{60}(46.67,37.33)(48.50,30.33)(46.50,22.67)
\bezier{80}(51.50,39.17)(55.50,30.17)(51.50,21.00)
\put(62.00,23.00){\vector(-3,1){14.50}}
\put(62.00,23.00){\vector(-4,3){8.00}}
\put(63.00,23.00){\makebox(0,0)[lc]{$r=$ const}}
\end{picture}

\noindent
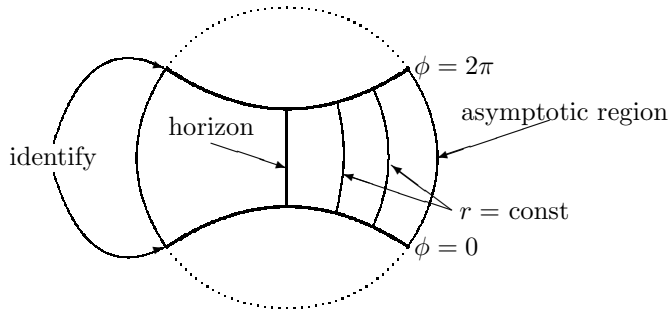
{\bf Fig.\ 1:} The BTZ black hole (region between the heavy lines), cut along
$\phi = 0$, as part of the Poincar\'e disk (dotted)

\bigskip

By means of a suitable isometry almost any $J=0$ black hole can be put
in the symmetrical position of Fig.\ 1. It is clear that the only
remaining difference between $J=0$ black holes is then the ``width" of the
region between the ultraparallels, which can be measured by the length
of the horizon, $2\pi r_+ = 2\pi\sqrt{M/|\Lambda|}$, or by the black
hole mass $M$. The exceptional case occurs
when the two geodesics are parallel. Since they meet at infinity, their
minimum distance, and the parameter $M$, is zero. Thus the horizon of this
``massless black hole" may be said to be at infinity, corresponding to
an infinitely long (and infinitely thin) throat.

\section{MBH Initial Values}

Instead of cutting the BTZ space along one radial geodesic ($\phi = 0$), we
can cut it along $\phi = 0$ and $\phi = \pi$. The resulting ``strips"
on the Poincar\'e disk are congruent. We can imagine them laid one on
top of the other, and sewn together at the radial boundaries, thus
restoring the BTZ geometry. (The identification is smooth because the
extrinsic and intrinsic geometries of these boundaries agree.) We denote
by {\em doubling} this procedure whereby two congruent copies of a region
are identified along totally geodesic boundaries.

This suggests a simple way to construct MBH geometries: Take any set of
mutually ultraparallel geodesics that bound a region in $H$, such as the
thick arcs in Fig.\ 2a, and double it (Fig.\ 2b).
The Figure has three-fold
symmetry, so the three black holes have equal masses. In general the masses
can have arbitrary values, including zero. (If all three masses are zero,
Fig.\ 2a is an ``ideal triangle" with vertices on the boundary of the
Poincar\'e disk.) If it is understood that any figure is to be doubled,
we can take a picture like Fig.\ 2a as a representation of the MBH geometry.
We shall call such a picture, showing the $n$ ultraparallels, a {\em diagram}
for the $n$-BH geometry.
The unique orthogonal geodesics to adjacent ultraparallel geodesics of a
diagram determine the horizon between those geodesics.

\bigskip
\unitlength 1.00mm
\linethickness{0.4pt}
\begin{picture}(152.00,58.40)(30,0)
\thicklines
\bezier{144}(24.50,42.50)(37.17,30.00)(24.50,17.50)
\bezier{140}(37.00,49.75)(41.50,32.67)(58.75,36.83)
\bezier{140}(37.00,10.33)(41.50,27.33)(58.75,23.00)
\thinlines
\bezier{64}(24.50,42.50)(29.00,48.50)(37.00,49.75)
\bezier{64}(58.75,37.17)(61.33,30.00)(58.75,23.00)
\bezier{64}(24.50,17.50)(29.00,11.50)(37.00,10.33)
\bezier{40}(37.00,49.75)(53.17,50.83)(58.75,37.17)
\bezier{40}(24.50,42.50)(15.00,30.00)(24.50,17.50)
\bezier{40}(37.00,10.33)(53.17,9.17)(58.75,23.00)
\bezier{15}(53.20,45.33)(49.60,41.87)(47.73,36.67)
\bezier{70}(47.73,36.67)(45.87,30.00)(47.87,23.33)
\bezier{15}(47.87,23.33)(49.47,18.00)(53.20,14.67)
\thicklines
\bezier{144}(100.91,30.01)(114.88,24.00)(100.91,17.99)
\bezier{140}(114.69,33.50)(119.65,25.28)(138.67,27.29)
\bezier{140}(114.69,14.54)(119.65,22.72)(138.67,20.63)
\thinlines
\bezier{64}(100.91,30.01)(105.87,32.90)(114.69,33.50)
\bezier{64}(138.67,27.45)(141.52,24.00)(138.67,20.63)
\bezier{64}(100.91,17.99)(105.87,15.10)(114.69,14.54)
\bezier{40}(114.69,33.50)(132.52,34.02)(138.67,27.45)
\bezier{40}(100.91,30.01)(90.44,24.00)(100.91,17.99)
\bezier{40}(114.69,14.54)(132.52,13.98)(138.67,20.63)
\bezier{15}(132.55,31.37)(128.58,29.71)(126.52,27.21)
\bezier{64}(126.52,27.21)(124.47,24.00)(126.68,20.79)
\bezier{15}(126.68,20.79)(128.44,18.23)(132.55,16.63)
\put(100.00,36.00){\line(-2,-3){16.00}}
\thicklines
\put(84.00,12.00){\line(1,0){68.00}}
\thinlines
\put(152.00,12.00){\line(-2,3){16.00}}
\put(136.00,36.00){\line(-1,0){36.00}}
\thicklines
\bezier{144}(100.91,53.71)(114.88,49.00)(100.91,44.29)
\bezier{140}(114.69,56.44)(119.65,50.01)(138.67,51.57)
\bezier{140}(114.69,41.59)(119.65,47.99)(138.67,46.36)
\thinlines
\bezier{64}(100.91,53.71)(105.87,55.97)(114.69,56.44)
\bezier{64}(138.67,51.70)(141.52,49.00)(138.67,46.36)
\bezier{64}(100.91,44.29)(105.87,42.03)(114.69,41.59)
\bezier{40}(114.69,56.44)(132.52,56.85)(138.67,51.70)
\bezier{40}(100.91,53.71)(90.44,49.00)(100.91,44.29)
\bezier{40}(114.69,41.59)(132.52,41.15)(138.67,46.36)
\bezier{15}(132.55,54.78)(128.58,53.47)(126.52,51.51)
\bezier{64}(126.52,51.51)(124.47,49.00)(126.68,46.49)
\bezier{15}(126.68,46.49)(128.44,44.48)(132.55,43.22)
\multiput(100.00,58.40)(-0.12,-0.14){134}{\line(0,-1){0.14}}
\thicklines
\put(84.00,39.60){\line(1,0){68.00}}
\thinlines
\multiput(152.00,39.60)(-0.12,0.14){134}{\line(0,1){0.14}}
\put(136.00,58.40){\line(-1,0){36.00}}
\put(87.00,34.00){\vector(2,-1){21.00}}
\put(87.00,38.00){\vector(2,1){21.00}}
\put(87.00,36.00){\makebox(0,0)[cc]{identify}}
\put(40.00,1.00){\makebox(0,0)[cc]{(a)}}
\put(120.00,1.00){\makebox(0,0)[cc]{(b)}}
\put(39.00,32.00){\vector(3,-1){8.00}}
\put(39.00,33.00){\makebox(0,0)[cb]{horizon}}
\end{picture}

\medskip \noindent
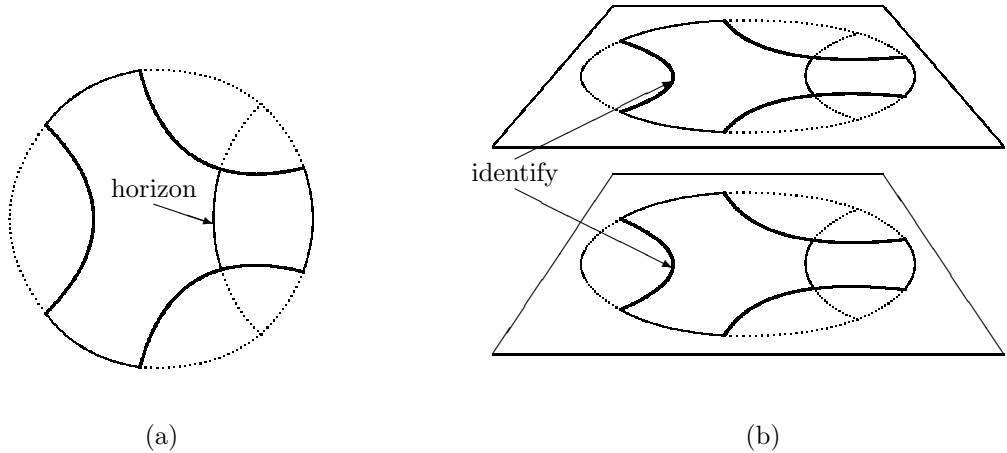
{\bf Fig.\ 2:} Construction of a three-black-hole initial geometry by
doubling a region bounded by three geodesics in the Poincar\'e disk.

\noindent
(a) Half of the initial geometry is represented by the region bounded by the
thick circular arcs.  The horizon of one black hole region
(minimal geodesic between the two geodesics on the right) is shown. The other
two horizons can be obtained by 120$^o$ rotations.

\noindent
(b) Two disks are placed one above the other, and the thick
boundaries are identified vertically, as shown explicitly for the boundary
on the left. The result is an initial state with three asymptotically adS
regions and three horizons.

\vskip 1cm

The region outside of each horizon is isometric to that of a BTZ black
hole --- one needs only to eliminate all but the two geodesics bounding
that region in order to obtain that ``single" BTZ black hole initial state.
Thus the exterior regions have all the properties of single black holes,
in particular the mass of each region is well defined. The interior region,
between the horizons, is a new type of initial value for 2+1 Einstein
theory. It  can be interpreted as an $n$-black-hole universe that is closed
except for the throats of the black holes (with $n=3$ in the example
above). The diagram of the interior region is a polygon with $2n$ sides that
meet at right angles. We call this the {\em polygon} of the diagram (Fig.\ 3a
gives an example). Alternate sides of the polygon represent the lengths of the
black hole horizons, and the distances between adjacent horizons.

Because the side lengths of a closed $90^\circ$ polygon cannot be assigned
arbitrarily, there are three relation between the masses and distances of a
MBH geometry so constructed. If we give an orientation to the polygon we can
regard the sides as
a sequence of transvections, or a sequence of boosts $\{L_i\}$ in the
Minkowski space embedding. The relation then demands that successive
application of all these boosts yield unity, $\prod_{i=1}^n L_i = 1\!\!1$.

If the lengths (or mass parameters) of two horizons are equal
we can identify two such horizons, rather than attaching the asymptotically
adS region on the other side. Thus one can also construct truly closed
``wormhole" universes. (This is just one set of the many identifications
possible in $H$ to form closed spaces.) However, the initial data as
constructed by doubling
are not the most general $J=0$ MBH geometries. For
example, we can cut the four-hole diagram of Fig.\ 3a along the dotted circle
and re-attach after turning through some angle (Fig.\ 3b). It is conjectured
that the general $J=0$ MBH geometry can be obtained from a diagram type by
rotations along such closed geodesics.

\unitlength 1.10mm
\linethickness{0.4pt}
\begin{picture}(110.83,50.00)(10,5)
\bezier{13}(10.00,30.00)(10.00,36.00)(14.00,42.00)
\bezier{20}(14.00,42.00)(20.00,50.00)(30.00,50.00)
\bezier{13}(50.00,30.00)(50.00,36.00)(46.00,42.00)
\bezier{20}(46.00,42.00)(40.00,50.00)(30.00,50.00)
\bezier{13}(10.00,30.00)(10.00,24.00)(14.00,18.00)
\bezier{20}(14.00,18.00)(20.00,10.00)(30.00,10.00)
\bezier{13}(50.00,30.00)(50.00,24.00)(46.00,18.00)
\bezier{20}(46.00,18.00)(40.00,10.00)(30.00,10.00)
\bezier{6}(10.33,33.67)(13.50,33.00)(16.67,34.83)
\thicklines\bezier{20}(16.67,34.83)(18.50,36.17)(19.83,38.67)
\thinlines\bezier{9}(19.83,38.67)(21.67,42.83)(19.00,46.83)
\bezier{40}(14.00,41.83)(16.50,40.00)(19.83,38.67)
\thicklines\bezier{88}(19.83,38.67)(30.00,34.67)(40.17,38.67)
\thinlines\bezier{40}(40.17,38.67)(43.83,40.00)(46.17,42.00)
\bezier{36}(12.00,21.00)(14.83,22.17)(16.50,25.17)
\thicklines\bezier{44}(16.50,25.17)(19.17,30.00)(16.50,34.67)
\thinlines\bezier{36}(16.50,34.67)(15.00,37.50)(12.17,39.00)
\bezier{7}(49.67,33.67)(46.50,33.00)(43.33,34.83)
\thicklines\bezier{20}(43.33,34.83)(41.50,36.17)(40.17,38.67)
\thinlines\bezier{36}(48.00,21.00)(45.17,22.17)(43.50,25.17)
\thicklines\bezier{44}(43.50,25.17)(40.83,30.00)(43.50,34.67)
\thinlines\bezier{36}(43.50,34.67)(45.00,37.50)(47.83,39.00)
\bezier{9}(40.17,38.67)(38.33,42.83)(41.00,46.83)
\bezier{40}(14.00,18.17)(16.50,20.00)(19.83,21.33)
\thicklines\bezier{88}(19.83,21.33)(30.00,25.33)(40.17,21.33)
\thinlines\bezier{40}(40.17,21.33)(43.83,20.00)(46.17,18.00)
\bezier{7}(10.33,26.33)(13.50,27.00)(16.67,25.17)
\thicklines\bezier{20}(16.67,25.17)(18.50,23.83)(19.83,21.33)
\thinlines\bezier{9}(19.83,21.33)(21.67,17.17)(19.00,13.17)
\bezier{7}(49.67,26.33)(46.50,27.00)(43.33,25.17)
\thicklines\bezier{20}(43.33,25.17)(41.50,23.83)(40.17,21.33)
\thinlines\bezier{9}(40.17,21.33)(38.33,17.17)(41.00,13.17)
\thicklines\bezier{20}(30.00,36.50)(30.00,30.00)(30.00,23.50)
\bezier{20}(71.67,27.17)(71.17,30.83)(72.83,31.00)
\bezier{36}(72.83,31.00)(75.83,29.83)(77.00,24.00)
\bezier{24}(77.00,24.00)(77.17,20.50)(74.83,21.50)
\bezier{28}(74.83,21.50)(72.50,23.67)(71.67,27.00)
\bezier{36}(80.50,36.50)(80.00,40.50)(84.00,41.17)
\bezier{32}(84.00,41.17)(87.50,42.00)(89.17,38.00)
\bezier{132}(76.50,21.50)(93.50,28.67)(101.00,16.00)
\bezier{48}(80.50,36.50)(80.67,32.67)(72.83,31.00)
\bezier{28}(98.67,42.67)(97.83,40.50)(102.50,38.33)
\bezier{32}(102.50,38.33)(107.83,36.33)(109.33,37.50)
\bezier{24}(109.33,37.50)(109.50,39.33)(105.83,41.67)
\bezier{32}(105.83,41.67)(101.00,43.83)(98.83,42.67)
\bezier{88}(89.17,38.00)(93.17,30.33)(98.50,42.50)
\bezier{28}(91.50,23.67)(89.33,24.50)(88.67,28.67)
\bezier{32}(88.67,28.67)(89.33,34.17)(92.17,35.00)
\bezier{72}(110.83,20.67)(106.83,29.00)(109.33,37.50)
\bezier{28}(101.00,16.00)(103.17,14.67)(107.33,16.50)
\bezier{24}(107.33,16.50)(110.67,18.17)(110.83,20.67)
\put(30.00,3.00){\makebox(0,0)[cc]{(a)}}
\put(92.00,3.00){\makebox(0,0)[cc]{(b)}}
\end{picture}

\vskip 0.5cm
\noindent
{\bf Fig.\ 3:} A diagram that allows rotation.\\
(a) A diagram for four black holes. Its polygon consists of the thick arcs.
The thick dotted line becomes a circle after the identifications.
The resulting geometry is to be cut along the thick dotted line, rotated by
some angle, and reassembled.\\
(b) 3D picture to give an idea of the result (cut off at the horizons).

\bigskip
The black hole or wormhole universes are related by a curious duality.
The dual universe is obtained by exchanging the roles of the horizons
and of the identification lines in the polygon. (A polygon with pairwise
equal sides is its own dual.)

\section{Time Development}

Because the initial data of each exterior region are the same as that of some
single BTZ black hole, the exterior time development will also be the same.
The Killing vector of that development can be extended even beyond the
horizon, but it does not extend to a global
symmetry. To discuss the global time development it is therefore more
appropriate to choose a constant lapse, $N = 1,\, N_i = 0$ (``time orthogonal"
development). With this choice
the diagram's identification lines develop into totally geodesic timelike
surfaces. These can therefore still be smoothly identified in the
three-dimensional time development.

Locally the time orthogonal development is the same as that of adS space:
successive spacelike surfaces have increasingly negative intrinsic curvature,
and acquire increasing, spatially constant, extrinsic curvature (with normals
converging in the advancing time direction). Globally, then, the intrinsic
geometry of the time surfaces is characterized by the same diagram as the
initial surface, but corresponding to larger negative curvature. The total
``volume" of the interior region (the area of the polygon)
decreases, as does the distance between black hole horizons. Physically
the interior MBH universe collapses, and the relative motion of the black
holes is similar to that of test particles in a background adS space.
(The same behavior was found in 3+1 dimensional black hole universes
\cite{KT,BHKT}.)

After a finite time the area tends to zero and the curvatures of the time
slices become infinite. In the adS universe this is not a spacetime
singularity, but in the BTZ and MBH geometry there is a singularity of
the Misner \cite{MS} type, where the spacetime fails to be Hausdorff
\cite{BHTZ}.
Physically we can say that, in these time-orthogonal coordinates, each
black hole horizon as well as the entire interior collapse simultaneously.
Just before the collapse the spacelike geometry looks like $n$ thin horns
that flare out at infinity and are connected on their thin ends. In this
sense the singularity consists of $n$ lines meeting at a point. (This point
is the endpoint of the unstable geodesic between the black holes that avoids
falling into any of them.)

Because there is no global Killing vector that moves the initial surface
in the MBH spacetimes we constructed, this initial surface is unique. In
this respect the MBH spacetimes differ from the BTZ black holes, where
the time-symmetric initial surfaces can be moved by the timelike
Killing vector.

Instead of generating MBH spacetimes from initial values and their time
development, we can also obtain them from an identification of the
3-dimensional adS spacetime. We still choose an initial time-symmetric
spacelike surface, and a diagram in it. We construct the totally geodesic
timelike surfaces ({\em identification surfaces}) normal to the initial
surface and intersecting it along a
diagram geodesic. We then double the spacetime between these surfaces by
sewing along the surfaces (a three-dimensional version of the procedure of
Fig.\ 2b). The identification surfaces will intersect somewhere in the
future and past of the initial surface.
This causes the non-Hausdorff singularity mentioned above.

\section{MBH with Angular Momentum}

MBHs with angular momentum can be constructed by a similar identification of
three-dimensional regions of adS space. We imitate the construction of the
BTZ ``single" black hole \cite{BTZ,C}. We cut this geometry along $\phi=0$
and embed the resulting three-dimensional ``slab" in three-dimensional adS
space. The line $r = r_+$ is an extremum of distance between the
identification surfaces $\phi = 0$ and $\phi = 2\pi$. We call it the
{\em horizon} even in the three-dimensional context. We replace the
identification surfaces by new,
totally geodesic identification surfaces that are generated by all geodesics
orthogonal to the horizon. This transforms the general metric (\ref{BTZ})
with $J \neq 0$ into one with $J=0$, mass $\mu$ and coordinates $\rho,
\varphi, \tau$ (and a different identification between
$\varphi = 0$ and $\varphi = 2\pi$):
\begin{eqnarray}\label{TG}
\mu = |\Lambda|r_+^2 &\qquad&
{\rho^2\over r_+^2} = {r^2-r_-^2 \over r_+^2 -r_-^2} \nonumber\\
\varphi = \phi - {r_-\sqrt{|\Lambda|}\over r_+}t &\qquad&
\tau = t - {r_-\over r_+\sqrt{|\Lambda|}}\phi
\end{eqnarray}

Consider the transvection whose invariant geodesic is the horizon. It maps
one new identification surface into the other, as for the case $J=0$.
The BTZ case $J \neq 0$ is different because the identification is not simply
given by this transvection. Rather, there is an additional ``twist," a boost
about the horizon. Let the boost take place in the surface $\varphi=2\pi$.
According to Eq (\ref{TG}) it can be described in
that surface  as a time translation by $2\pi r_-/r_+\sqrt{|\Lambda|}$.
Thus the BTZ black hole with angular momentum can be obtained from a $J=0$ BTZ
black hole with the same horizon length by shifting the $J=0$ BTZ time by
this amount on the second identification surface.

To obtain an analogous MBH solution we must find several identification
surfaces
and horizons that can be consistently doubled in the presence of twists,
and in three-dimensional adS spacetime.
These surfaces and horizons can again
be specified by a series of isometries of adS spacetime. In addition to
transvections that move one surface into the next, and transvections along the
geodesics adjoining adjacent horizons, there should now also be boosts about
the horizons. Consistency demands that the product of all these
transformations be unity, a set of six conditions among the masses, distances,
and angular momenta.

We can again associate a diagram with these transformations,
consisting of mutually orthogonal geodesics that represent alternately the
connections between horizons and the horizons themselves. The geodesics
are the invariant geodesics of the transvections; the boosts are
measured by the amount by which the directions of the connecting geodesics
at the two ends of a horizon fail to be parallel, as judged by parallel
transport along the horizon. Thus such a diagram cannot lie in a plane
(totally geodesic spacelike) surface and is therefore harder to visualize.
The identification surfaces are then generated by
all the geodesics orthogonal to the horizons at their ends.
(The same surface is generated from either of the horizons ending on it.)
Finally the spacetime region bounded by these surfaces
is doubled by identifying along the surfaces with an second, similar region.
In order that the twist not cancel when going all the way around a horizon in
the doubled spacetime, the second region should be identically constructed but
with twists in the opposite directions.

If there are three identification surfaces, the horizon lines always lie in
the plane through the surfaces' three centers, so the diagram is planar
and corresponds to $J=0$. For nonzero angular momenta we therefore need
at least four black holes in our construction. To show that such diagrams
and identification surfaces exist we give one example. Consider the
diagram of Fig.\ 3a, and the four identification surfaces generated by all
geodesics orthogonal to the horizons at their endpoints. Boost the part
to the right of the thick dotted line by a Lorentz transformation that has
this dotted line as its axis. The left identification surface and the two
horizons on the left are not affected
by this boost of the right part, and the top and bottom identification
surfaces are invariant under this boost; however, the right identification
surface and the two right horizons move. Likewise, the geodesics connecting
the horizons on the left with those on the right move. As a result there
is now a twist between the long and the short connecting geodesics, the
diagram is no longer planar, and there is angular momentum associated with
each horizon.

\section{Conclusion}

We have seen that it is possible to construct out of pieces of adS spacetime
a spacetime that has many asymptotically adS regions containing many horizons.
Each of these regions is isometric to the corresponding region of a BTZ black
hole. It is therefore appropriate to regard such spacetimes as MBH spacetimes.
For the case of zero angular momentum our construction can be characterized
by a polygon whose sides represent the distances and masses involved. The
closure condition yields three relations between these parameters; but other,
``non-polygonal" arrangements can also be constructed. In the case $J \neq 0$
we get six rather less transparent conditions between masses, distances,
and angular momenta; but more general MBH configurations, not obtained by
simple doubling, presumably exist.

\begin{flushleft}
\large\bf Added Note
\end{flushleft}
After this paper was finished I heard from Dr. Alan Steif of UC Davis that he
has also found many of the results of the present paper's sections 2 and 3.

\end{document}